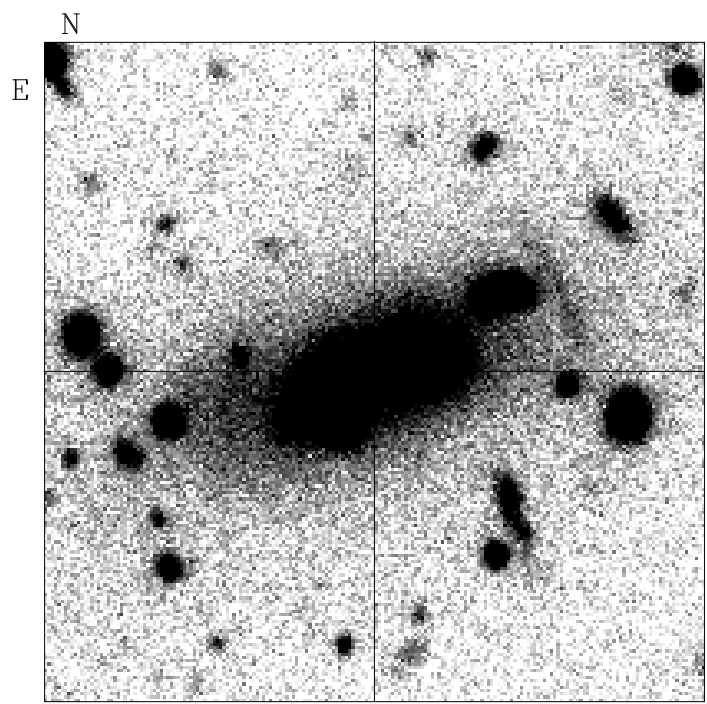

# Discovery of a Large Gravitational Arc in the X-ray Cluster A2280*

I. M. Gioia[1,2], J. P. Henry[1], G. A. Luppino[1], D. I. Clowe[1], H. Böhringer[3], U. G. Briel[3], W. Voges[3], J.P. Huchra[4] and H. MacGillivray[5]

[1] Institute for Astronomy, 2680 Woodlawn Drive, Honolulu, HI 96822, USA
[2] Istituto di Radioastronomia del CNR, Via Gobetti 101, I–40129 Bologna, Italy
[3] Max-Planck-Institut für Extraterrestrische Physik D–85748 Garching bei München, Federal Republic of Germany
[4] Harvard–Smithsonian Center for Astrophysics, 60 Garden Street, Cambridge, MA 02138, USA
[5] Royal Observatory Edinburgh, Edinburgh, United Kingdom

**Abstract.** This *Letter* presents the serendipitous discovery of a large arc in an X-ray selected cluster detected in the *ROSAT* North Ecliptic Pole (*NEP*) survey. The cluster, associated with Abell 2280, is identified as the optical counterpart of the X-ray source $RXJ1743.5+6341$. This object is a medium–distant ($z=0.326$) and luminous ($L_{0.5-2kev} = 5.06 \times 10^{44}$ ergs s$^{-1}$) cluster dominated by a central bright galaxy. The arc is located $\sim 14''$ to the North–West of the cD and it is detected in the R and I bands but not in the B band. Photometric and spectroscopic observations of the cluster and of the arc are presented. Other similar discoveries in the course of imaging and spectroscopic surveys of *ROSAT* are expected in the future.

## 1. Introduction

Evidence for gravitational lensing in the form of giant arcs and faint arclets in rich clusters of galaxies is mounting thanks to the development of new observational imaging and spectroscopic techniques. Observations have moved in two directions: studies of individual arcs to investigate the distribution of dark matter in the central cores of the clusters, and studies of large samples to obtain information on the background source population of faint galaxies. In a recent review, Fort and Mellier (1994) provide an excellent update of the present status of gravitational arc surveys. Until recently most of the examples of lensing have been found in optically selected rich clusters. Some studies have relied on radio selected samples, and one of these has found the most distant cluster showing an arc (the arc associated with 3C220.1 at $z=0.62$; Dickinson 1992). However, the preferred method remains X-ray selection as demonstrated by the high frequency of lensing resulting from a study of the most X-ray luminous clusters from the *Einstein Observatory* Extended Medium Sensitivity Survey (EMSS; Gioia *et al.* 1990a; Stocke *et al.* 1991). The high X-ray luminosity is an indicator of a deep potential well characteristic of a massive cluster, an environment more likely to exhibit the lensing phenomenon. The preliminary results in Luppino and Gioia (1992), Luppino *et al.* (1993), Hammer *et al.* (1993), Le Fèvre *et al.* (1994) and the more recent arc survey of the complete sample by Luppino *et al.* (1995) confirm this hypothesis.

The arc reported in this *Letter* was serendipitously discovered during the optical identification of all the sources detected in the *NEP* region (Böhringer *et al.* 1991) of the *ROSAT* All-Sky Survey (RASS) (Trümper 1992), a $9° \times 9°$ region centered on the *NEP* (Gioia 1993, Henry *et al.* 1994), where the *RASS* exposure is the deepest. The X-ray source is positionally coincident with the cluster Abell 2280. The aim of the *NEP* survey program is to extract a substantial and well defined cluster sample, at redshifts extending potentially to $z \sim 1$. The resulting cluster catalog will be used to investigate the evolution of cluster X-ray and optical properties and for studies of large scale structure. The *RASS* has the potential to discover new

---

gravitational lensing phenomena which are more likely to be exhibited by rich and massive clusters found through direct X-ray selection. The recent reports of two bright arcs in $RXJ1347.5-1145$ (Schindler et al. 1995), of the red arc in cluster A2104 (Pierre et al. 1994) and of the giant arc in S295 (Edge et al. 1994), are all serendipitous discoveries in the course of optical follow-up observations of ROSAT clusters. This Letter presents X-ray and optical observations of the NEP cluster $RXJ1743.5+6341$ where such an arc was detected. Throughout, $H_0 = 50$ km s$^{-1}$ Mpc$^{-1}$ and $q_0 = 0.5$ are assumed.

## 2. Observations and Data Reduction

### 2.1. X-ray Observations

The X-ray source $RXJ1743.5+6341$ was detected at $17^h43^m28^s.2+63°41'45'' \pm 12''$ (J2000.0) (90% confidence error circle radius) in a region of the NEP survey which was exposed for 3086 seconds. The net count rate within a 6' radius circle was $6.59 \pm 0.59 \times 10^{-2}$ cnts s$^{-1}$ in the 0.4–2.4 keV band. The use of a Raymond spectrum with kT=5 keV, abundances 30% solar, a neutral hydrogen column density of $3.04 \times 10^{20}$ (Stark et al. 1992) and the redshift of the cluster, z=0.326, resulted in a conversion factor of 1 cnt s$^{-1}$ (0.4–2.4 keV) = $1.49 \times 10^{-11}$ ergs cm$^{-2}$ s$^{-1}$ (0.5–2 keV band). The X-ray flux adopting this conversion is $F_{0.5-2keV} = 9.8 \pm 0.9 \times 10^{-13}$ ergs cm$^{-2}$ s$^{-1}$ resulting in an X-ray luminosity in the same energy band of $L_x = 5.06 \times 10^{44}$ ergs s$^{-1}$. The total net counts (about 200) were insufficient to fit a meaningful spectrum. The X-ray contour map of the source, obtained by Gaussian smoothing the data with a $\sigma = 36''$ is shown in Fig. 1. The small number of photons and the resolution of the PSPC in the survey mode prevent any determination of the morphology (i.e. ellipticity) of the x-ray source at present.

### 2.2. Optical Observations

As part of the NEP identification process the University of Hawaii 2.2m telescope on Mauna Kea is used to obtain B and I band CCD images for the 629 X-ray sources detected in the 9° × 9° region centered on the NEP. The aim of the imaging is to separate sources associated with distant clusters from other sources using colors and morphology. The positional uncertainty of the X-ray source is also used as a discriminator. A stellar object falling within few arcseconds ($\leq 15''$) from the X-ray position is almost certainly the X-ray emitter. Within that positional uncertainty faint blue stellar objects or galaxies not obviously associated with a cluster are almost inevitably confirmed with follow-up spectroscopy to have characteristics of active galactic nuclei (AGN's). Nearby clusters (z$\lesssim$0.1) are usually extended in X-rays and identification can be made by inspection of Palomar Schmidt plates, obtained in two colors specifically for this project, and already digitized by the Royal Observatory Edinburgh. Distant cluster candidates are targeted through the analysis of the two-color CCD images and then observed spectroscopically using both the UH 2.2m and the CFHT 3.6m. $RXJ1743.5+6341$ was first imaged with the UH 2.2m telescope. A spectroscopic observation to measure the redshift of the cD galaxy was obtained right after the images were acquired. Multi-Object Slit (MOS) spectroscopy for the cluster galaxies was performed using the CFHT and the MARLIN focal reducer. The wavelength resolution is 4.1 Å pixel$^{-1}$ in the spectral range $\lambda\lambda$ 4500–8500 Å. The journal of observations is given in Table 1, along with the measured redshifts and magnitudes (within a 20 kpc radius).

**Fig. 1.** X-ray isointensity contour map of RX J1743.5+6341 in sky pixels coordinates. North is up and East to the left. The X-ray emission is centered at $17^h43^m28^s.2+63°41'45''$ Each tic mark corresponds to 50''. Contour levels are in steps of $0.35 \times 10^{-3}$ cnts $(24'' \times 24'')^{-2}$s$^{-1}$ with the first contour at $1.05 \times 10^{-3}$ and the last contour at $5.25 \times 10^{-3}$ cnts $(24'' \times 24'')^{-2}$s$^{-1}$ above a background level of $6.5 \times 10^{-5}$ cnts $(24'' \times 24'')^{-2}$s$^{-1}$.

An apparent arc was visible in the I band image and follow-up observations to confirm it were obtained in R band. The conditions were photometric with a seeing of 0.8''–0.9''. The individual bias-subtracted and flattened images were shifted into registration and stacked while being cleaned of cosmic rays. The new faint standards of Landolt (1992) were used for photometric calibration. A sub-array region of the CCD R frame is shown in Fig. 2.

## 3. Results and Discussion

An appropriate analysis of the X-ray source extent of $RXJ1743.5+6341$ is prevented by the low statistics. The peak of the X-ray emission is coincident within 15″ with the brightest cluster galaxy of Abell 2280, which is classified by Abell, Corwin and Olowin (1989) as B–M type II–III, R=0, D=6. CFHT MOS spectroscopy data are available for 13 objects. Seven galaxies are compatible with being cluster members. One galaxy is a foreground object with Balmer emission lines at $z=0.205$, five spectra have inadequate signal-to-noise ratio for a reliable redshift determination. Table 1 gives the offsets in position between each galaxy and the cD galaxy in arcseconds. Negative offsets indicate direction west and south of the cD. The mean redshift of the cluster is $z=0.326 \pm 0.002$ which makes $RXJ1743.5+6341$ moderately luminous at $L_x=5.06\times10^{44}$ ergs s$^{-1}$ (0.5–2 keV). A tentative determination of the velocity dispersion (according to the formula of Danese et al. 1980) gives $948^{+516}_{-285}$ km s$^{-1}$.

**Table 1.** Journal of Observations

| Date(UT) | Telescope | CCD | Observations |
|---|---|---|---|
| June 13, '91 | UH 2.2m | Tek 1024 | 900s B, 600s I |
| June 13, '91 | UH 2.2m | Tek 1024 | 1800s, longslit |
| June 26, '92 | CFHT 3.6m | SAIC1 2048 | 5400s, multislit |
| May 18, '93 | UH 2.2m | Tek 2048 | 2400s R |

| | Results | | |
|---|---|---|---|
| galaxy # | $z \pm \Delta z$ | Offsets in " | B, R, I |
| galaxy 1 | 0.331±0.002 | 0, 0 | 20.6, 17.9, 17.2 |
| galaxy 2 | 0.321±0.003 | −52, −10 | 23.4, 20.5, 20.1 |
| galaxy 3 | 0.332±0.003 | −36, +18 | 21.9, 19.4, 18.9 |
| galaxy 4 | 0.322±0.001 | −24, −39 | 22.6, 19.9, 19.3 |
| galaxy 5 | 0.324±0.003 | +42, +50 | 21.9, 19.3, 18.6 |
| galaxy 6 | 0.321±0.001 | +94, +39 | 20.9, 18.7, 18.2 |
| galaxy 7 | 0.329±0.001 | +124, +23 | 21.3, 18.5, 17.9 |
| galaxy 8 | 0.205±0.002 | −74, +23 | 20.2, 18.4, 17.9 |

The presence of the radio source $8C1743+637$ in the region of Abell 2280 prompted an optical study by Lacy et al. (1993) of the cluster, although the radio emission was later identified with a galaxy at a much higher redshift than the cluster (the radio source is outside the region presented in Fig. 2). Spectra of the cD galaxy and of a foreground spiral at z = 0.084 have been presented in the literature (see Figures 1 and 3 by Lacy et al. 1993). Their measurement of the cD redshift (0.33±0.005) agrees very well with our determination.

Our 2400s R image shows a rather regular cluster with a bright cD (R=17.9 within a 20 kpc radius). The cD has a few companion galaxies embedded in a common envelope, similar to other X-ray selected clusters. Several cases of multiple-nuclei systems have been found in the EMSS (Gioia & Luppino 1994) and were interpreted (Annis 1994) as a possible evolutionary effect in the morphologies of the brightest cluster galaxies over look-back times corresponding to redshift $z=0.3$ or higher. In the CCD image the cluster appears richer than the Abell classification. The detection and classification package FOCAS (Valdes 1982) has been used to determine the parameter $N_{0.5}$ (Bahcall 1981), the number of bright galaxies (m≤ $m_3$+2) projected within 0.5 Mpc radius from the cD galaxy. The corrected counts after subtraction of estimated background of 4 galaxies according to Tyson (1988) gives $N_{0.5}=37$, which transforms to about 100 Abell counts, corresponding to an Abell richness 2 cluster. The $N_{0.5}$ value found is higher than the expected value from the Bahcall (1981) $N_{0.5}-\sigma$ correlation (37 versus $19^{+14}_{-7}$) when we use the measured value for the velocity dispersion. However, this result must remain tentative until a more precise velocity dispersion is obtained. According to the $L_x$–$N_{0.5}$ correlation for Abell clusters (see Fig. 3 in Bahcall 1980), Abell 2280 seems to be richer than expected from its X-ray luminosity.

**Fig. 2.** CCD R image of the central 44″ ×44″ region of $RX$ J1743.5+6341. North is up and East to the left. The linear scale at the redshift of the cluster is 5.8 kpc arcsecond$^{-1}$ corresponding to a side of 255×255 kpc in the figure. The large arc visible to the North-West is about 14″ from the cD. The cross marks the position of the cD galaxy center (x=0, y=0 in Table 1)

The arc visible in Fig. 2 lies 13.8″ North–West from the center of the cD galaxy. The radius of curvature is consistent with the distance from the center of the arc to the center of the cluster, i.e. the center of curvature coincides with the cD. The total length of the arc measured on the higher resolution R frame is 7.9″ and consists of 2 pieces. The NE part is 4.6″ long and appears to be unresolved in width in $0.8''-0.9''$ seeing. A more diffuse section is visible at the SW end of the arc, with variation in width up to $2''$. Accurate photometry of the arc is difficult given its proximity to the cD halo. The arc total R magnitude was measured by integrating all the flux enclosed by the lowest isophote. We obtain a magnitude $m_R = 22.9 \pm 0.2$ and a surface brightness $\mu_R = 25.0$ mag arcsec$^{-2}$. To minimize contamination from the cD envelope light the sky background was computed from an equivalent region symmetric to the arc on the other side of the cD galaxy. The arc is not detected in a B image down to a surface brightness of $\mu_B = 27.3$ mag arcsec$^{-2}$, from which a B−R color index $>2.3$ can be derived. This limit is at the upper red end of the color range of other known arcs (0.5−2.2, Soucail, 1992), and it is similar to that of the arc found in the bright X-ray cluster A2104 (Pierre et al. 1994). The seeing in our B image was not very good ($\sim 1.4''$), however. In the I frame the arc has a magnitude $m_I = 21.1 \pm 0.1$ and a surface brightness of $\mu_I = 24.1$ mag arcsec$^{-2}$.

If we take the radius of curvature as the Einstein ring radius, $R_E$, we can give a crude estimate of the central mass projected inside $R_E = 79$ kpc. A spectroscopic measurement of the arc redshift is not available. Assuming as a reasonable guess the source to be at a redshift of 0.7, twice the distance of the cluster, a mass of the order of $6 \times 10^{13}$ $M_\odot$ is obtained. The mass drops to $4.7 \times 10^{13}$ $M_\odot$ if we move the source out to a redshift of $z=1$. These values are within the range of the total mass within the critical radius displayed by cluster-lens systems for which a modelling has been attempted using either giant arcs or arclets (see Table 2 in Fort and Mellier, 1994).

*Acknowledgements.* We thank the UH Time Allocation Committee for continuous allocation of UH 2.2m and CFHT time for the *NEP* project. This work received partial financial support from NSF Grants AST–91199216, AST–9020680, NASA Grants NAG5–1752, NAG5–1880, and NAG5–2594 and BMFT. JHP and UGB received support from NATO.